\documentclass[twoside,twocolumn,9pt]{article}
\usepackage{extsizes}
\usepackage[super,sort&compress,comma]{natbib} 
\usepackage[version=3]{mhchem}
\usepackage[left=1.5cm, right=1.5cm, top=1.785cm, bottom=2.0cm]{geometry}
\usepackage{balance}
\usepackage{times,mathptmx}
\usepackage{sectsty}
\usepackage{graphicx} 
\usepackage{lastpage}
\usepackage[format=plain,justification=justified,singlelinecheck=false,font={stretch=1.125,small,sf},labelfont=bf,labelsep=space]{caption}
\usepackage{float}
\usepackage{fancyhdr}
\usepackage{fnpos}
\usepackage[english]{babel}
\usepackage{array}
\usepackage{droidsans}
\usepackage{charter}
\usepackage[T1]{fontenc}
\usepackage[usenames,dvipsnames]{xcolor}
\usepackage{setspace}
\usepackage[compact]{titlesec}
\usepackage{amsfonts, amsmath, amssymb, amsthm, array, eucal, mathtools, xfrac,
booktabs, array, natbib, graphicx, siunitx, color, subcaption, contour}
\contourlength{0.3pt} 
\contournumber{100}  
\usepackage{graphicx,tikz,pgfplots}

\pgfplotsset{
    colormap={parula}{
        rgb=(0.2081,0.1663,0.5292)
        rgb=(0.2116,0.1898,0.5777)
        rgb=(0.2123,0.2138,0.627)
        rgb=(0.2081,0.2386,0.6771)
        rgb=(0.1959,0.2645,0.7279)
        rgb=(0.1707,0.2919,0.7792)
        rgb=(0.1253,0.3242,0.8303)
        rgb=(0.0591,0.3598,0.8683)
        rgb=(0.0117,0.3875,0.882)
        rgb=(0.006,0.4086,0.8828)
        rgb=(0.0165,0.4266,0.8786)
        rgb=(0.0329,0.443,0.872)
        rgb=(0.0498,0.4586,0.8641)
        rgb=(0.0629,0.4737,0.8554)
        rgb=(0.0723,0.4887,0.8467)
        rgb=(0.0779,0.504,0.8384)
        rgb=(0.0793,0.52,0.8312)
        rgb=(0.0749,0.5375,0.8263)
        rgb=(0.0641,0.557,0.824)
        rgb=(0.0488,0.5772,0.8228)
        rgb=(0.0343,0.5966,0.8199)
        rgb=(0.0265,0.6137,0.8135)
        rgb=(0.0239,0.6287,0.8038)
        rgb=(0.0231,0.6418,0.7913)
        rgb=(0.0228,0.6535,0.7768)
        rgb=(0.0267,0.6642,0.7607)
        rgb=(0.0384,0.6743,0.7436)
        rgb=(0.059,0.6838,0.7254)
        rgb=(0.0843,0.6928,0.7062)
        rgb=(0.1133,0.7015,0.6859)
        rgb=(0.1453,0.7098,0.6646)
        rgb=(0.1801,0.7177,0.6424)
        rgb=(0.2178,0.725,0.6193)
        rgb=(0.2586,0.7317,0.5954)
        rgb=(0.3022,0.7376,0.5712)
        rgb=(0.3482,0.7424,0.5473)
        rgb=(0.3953,0.7459,0.5244)
        rgb=(0.442,0.7481,0.5033)
        rgb=(0.4871,0.7491,0.484)
        rgb=(0.53,0.7491,0.4661)
        rgb=(0.5709,0.7485,0.4494)
        rgb=(0.6099,0.7473,0.4337)
        rgb=(0.6473,0.7456,0.4188)
        rgb=(0.6834,0.7435,0.4044)
        rgb=(0.7184,0.7411,0.3905)
        rgb=(0.7525,0.7384,0.3768)
        rgb=(0.7858,0.7356,0.3633)
        rgb=(0.8185,0.7327,0.3498)
        rgb=(0.8507,0.7299,0.336)
        rgb=(0.8824,0.7274,0.3217)
        rgb=(0.9139,0.7258,0.3063)
        rgb=(0.945,0.7261,0.2886)
        rgb=(0.9739,0.7314,0.2666)
        rgb=(0.9938,0.7455,0.2403)
        rgb=(0.999,0.7653,0.2164)
        rgb=(0.9955,0.7861,0.1967)
        rgb=(0.988,0.8066,0.1794)
        rgb=(0.9789,0.8271,0.1633)
        rgb=(0.9697,0.8481,0.1475)
        rgb=(0.9626,0.8705,0.1309)
        rgb=(0.9589,0.8949,0.1132)
        rgb=(0.9598,0.9218,0.0948)
        rgb=(0.9661,0.9514,0.0755)
    }
}

\definecolor{cream}{RGB}{222,217,201}

\begin{document}

\pagestyle{fancy}
\thispagestyle{plain}
\fancypagestyle{plain}{

\fancyhead[C]{\vspace{1.1cm}\hrule\vspace{1.5cm}\hrule}

\renewcommand{\headrulewidth}{0pt}
}

\makeFNbottom
\makeatletter
\renewcommand\LARGE{\@setfontsize\LARGE{15pt}{17}}
\renewcommand\Large{\@setfontsize\Large{12pt}{14}}
\renewcommand\large{\@setfontsize\large{10pt}{12}}
\renewcommand\footnotesize{\@setfontsize\footnotesize{7pt}{10}}
\makeatother

\renewcommand{\thefootnote}{\fnsymbol{footnote}}
\renewcommand\footnoterule{\vspace*{1pt}%
\color{cream}\hrule width 3.5in height 0.4pt \color{black}\vspace*{5pt}} 
\setcounter{secnumdepth}{5}

\makeatletter 
\renewcommand\@biblabel[1]{#1}            
\renewcommand\@makefntext[1]%
{\noindent\makebox[0pt][r]{\@thefnmark\,}#1}
\makeatother 
\renewcommand{\figurename}{\small{Fig.}~}
\sectionfont{\sffamily\Large}
\subsectionfont{\normalsize}
\subsubsectionfont{\bf}
\setstretch{1.125} 
\setlength{\skip\footins}{0.8cm}
\setlength{\footnotesep}{0.25cm}
\setlength{\jot}{10pt}
\titlespacing*{\section}{0pt}{4pt}{4pt}
\titlespacing*{\subsection}{0pt}{15pt}{1pt}

\fancyfoot{}
\fancyhead{}
\renewcommand{\headrulewidth}{0pt} 
\renewcommand{\footrulewidth}{0pt}
\setlength{\arrayrulewidth}{1pt}
\setlength{\columnsep}{6.5mm}
\setlength\bibsep{1pt}

\makeatletter 
\newlength{\figrulesep} 
\setlength{\figrulesep}{0.5\textfloatsep} 

\newcommand{\topfigrule}{\vspace*{-1pt}%
\noindent{\color{cream}\rule[-\figrulesep]{\columnwidth}{1.5pt}} }

\newcommand{\botfigrule}{\vspace*{-2pt}%
\noindent{\color{cream}\rule[\figrulesep]{\columnwidth}{1.5pt}} }

\newcommand{\dblfigrule}{\vspace*{-1pt}%
\noindent{\color{cream}\rule[-\figrulesep]{\textwidth}{1.5pt}} }

\makeatother

\twocolumn[
  \begin{@twocolumnfalse}
\vspace{2.2cm}
\sffamily

 \noindent\LARGE{\textbf{Microscale Flow
Dynamics of Ribbons and Sheets}} \\

 \noindent\large{Thomas D. Montenegro-Johnson,\textit{$^{a,b,c}$} Lyndon
Koens\textit{$^{b}$} and Eric Lauga\textit{$^{b}$}} \\

 \noindent\normalsize{

Numerical study of the hydrodynamics of thin sheets and ribbons presents difficulties associated with resolving multiple length scales. To circumvent
these difficulties, asymptotic methods have been developed to describe the
dynamics of slender fibres and ribbons. However, such theories entail
restrictions on the shapes that can be studied, and often break down in regions
where standard boundary element methods are still impractical. In this paper we
develop a regularised stokeslet method for ribbons and sheets in order to
bridge the gap between asymptotic and boundary element methods. The method is
validated against the analytical solution for plate ellipsoids, as well as the
dynamics of ribbon helices and an experimental microswimmer. We then demonstrate
the versatility of this method by calculating the flow around a double helix,
and the swimming dynamics of a microscale ``magic carpet''.} \\

 \end{@twocolumnfalse} \vspace{0.6cm}

  ]

\renewcommand*\rmdefault{bch}\normalfont\upshape
\rmfamily
\section*{}
\vspace{-1cm}


\footnotetext{\textit{$^{a}$~School of Mathematics, University of Birmingham,
Edgbaston, Birmingham, B15 2TT}}
\footnotetext{\textit{$^{b}$~Department of Applied Mathematics and Theoretical
Physics, University of Cambridge, Centre for Mathematical Sciences, Wilberforce
Rd, Cambridge, CB3 0WA}}
\footnotetext{\textit{$^{c}$~Email for correspondance: t.d.johnson@bham.ac.uk}}



\section{Introduction}

At microscopic scales, fluid flow is governed by the Stokes flow equations:
inertialess, and kinematically reversible.~\citep{Kim2005} The dynamics of
microscale objects in Stokes flow is highly dependent upon their geometry; for
example, the drag anisotropy of slender rods can cause them to translate
at an angle to gravity when settling,~\citep{Guyon2001} and is crucial for the
swimming of microorganisms with flagellar filaments.~\citep{BECKER2003,Lauga2009} But for many systems of
medical or industrial significance, this geometric dependence can be difficult
to compute. 

\begin{figure*}[tb]
    \begin{center}
    \includegraphics{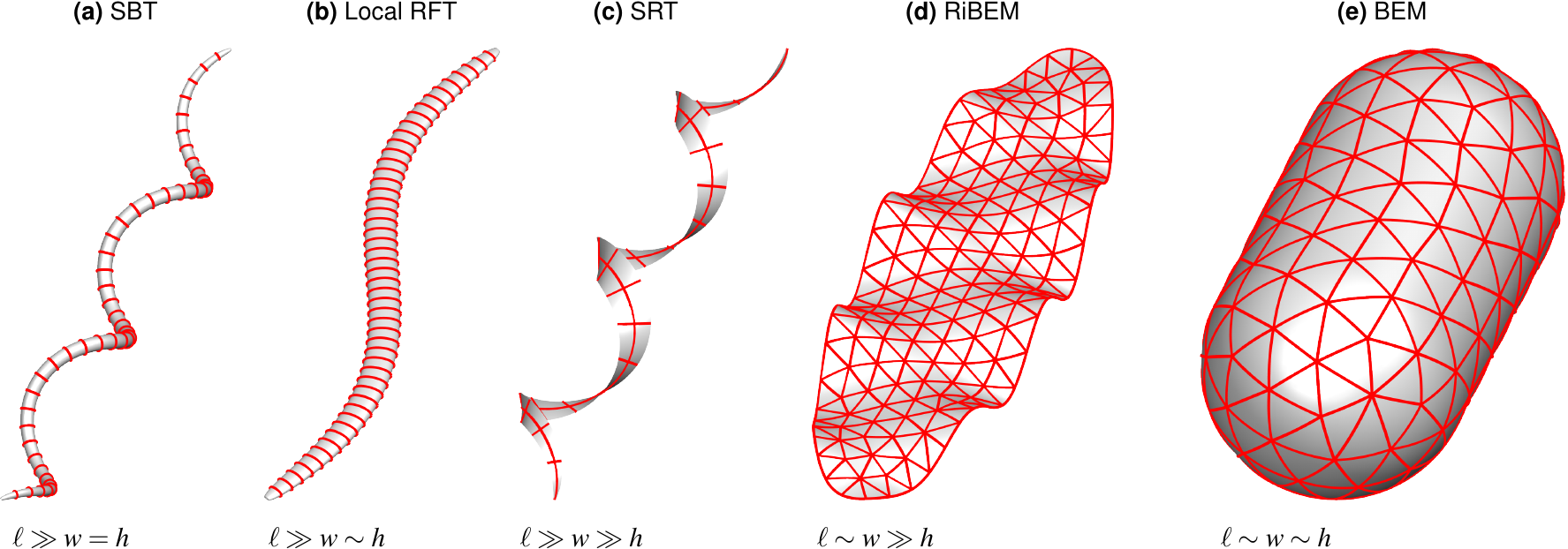}
    \end{center}
    \caption{Classes of shapes in Stokes flow. (a) Slender Body Theory (SBT) is
    an appropriate choice for long, slender bodies of circular cross-section.
    Examples are the flagella of bacteria or spermatozoa. (b) When this cross-section
is flattened slightly, but the width and thickness are still comparable, only
local drag theory (RFT) has so far been developed~\citep{Batchelor1970}; there is no non-local
equivalent to SBT in this case, but though inefficient,
boundary element methods may still be employed. An example of such a squashed
slender body is the trematode worm
\textit{Schistosoma mansoni}, responsible for Schistosomiasis. (c) Slender Ribbon
Theory (SRT) is appropriate when all three intrinsic length scales are separated $\ell
\gg w \gg h$, which can occur in artificial microswimmer design. (d) The current
study sits in the domain between SRT and the Boundary Element Method (BEM) (e).}
\label{fig:shape_classes}
\end{figure*}

The Boundary Element Method (BEM)~\citep{youngren1975stokes,Pozrikidis1992} is
commonly employed in studies of microscale
biofluiddynamics.~\citep{phan1987boundary,shum2010modelling,montenegro2014optimal,montenegro2016flow}
In BEM, the Stokes flow equations are transformed into an integral of Green's
functions over the domain boundaries.~\citep{Pozrikidis1992} These boundaries
are discretised or ``meshed'' into typically triangular elements
(Fig.~\ref{fig:shape_classes}e), and the boundary integral is then calculated
over each mesh element to form a linear system which can be solved. 

However, when there is a separation of length scales in the object being studied
(Fig.~\ref{fig:shape_classes}), as with for instance the slender flagellum of
bacteria or spermatozoa, resolving each scale with the boundary element method
can result in very large linear systems. Early work by
Hancock~\cite{hancock1953self} treated the slender flagellum of sea-urchin
spermatozoa as a line of point forces, from which local resistive force theories
were derived,~\citep{GRAY1955, Cox}, which could be generalised to any
cross-sectional shape (Fig.~\ref{fig:shape_classes}b)~\citep{Batchelor1970}.
However this local treatment required the cross-sectional length scale of the
filament, $w$, to be exponentially smaller than the centreline length, $\ell$
($1\gg1/\log(\ell/w)$), with small curvature. 

Building on this work, Slender Body Theory (SBT) was developed to determine the
flow from such filaments to algebraic accuracy.~\citep{1976,Johnson1979} These
models describe the hydrodynamics of a filament, with circular cross section, by
expanding the flow from a line of point forces placed along the filament's
centreline and has been used extensively to examine swimming and pumping in
microscale biological
systems.~\citep{Koens2014,Kim2004,Spagnolie2010,Spagnolie2011}

 Recently, slender-ribbon theory (SRT) was developed to accurately explore the
 microscale hydrodynamics of slender ribbons.~\citep{Koens2016, Koens2016a} This
 method expanded on the work of Johnson~\citep{Johnson1979} and so captures the
 non-local interactions of the shape and is accurate to order $w/\ell$. The
 slenderness condition in SRT is contingent on the assumption that all three
 internal length scales of the ribbon, the length, $\ell$, width, $w$, and
 thickness, $h$, are separated $\ell \gg w \gg h$
 (Fig.~\ref{fig:shape_classes}c), and the leading order flow is found by
 asymptotically expanding the flow from a plane of point forces representing the
 ribbon with respect to these length scales. The flow at the ribbon's surface is
 then given in terms of a line integral~\citep{Koens2016a} of unknown forces, as
 in SBT, which can be inverted to determine the force on the fluid. This
 technique allows the asymptotic exploration of a range of ribbon structures,
 such as artificial microswimmers comprising a magnetic head and a ribbon
 tail.~\citep{Zhang2009b, Zhang2009, Zhang2010}

While SRT captures the leading order behaviour of many ribbon shapes, it is
unsuitable for highly-twisted ribbons, or those with a very curved centreline.
Similarly, the asymptotic treatment of the ``slender-ribbon'' limit $\ell \gg w
\gg h$ prevents SRT from capturing the dynamics of
sheets~\citep{diller2014continuously} with $\ell \sim w \gg h$
(Fig.~\ref{fig:shape_classes}d), and causes the theory to break down before a
standard boundary element approach becomes practical.

In this paper we develop Ribbon-BEM (RiBEM), a method to bridge the gap between SRT and
BEM that is capable of solving the dynamics of both highly twisted and curved
ribbons as well as sheets. Inspired by previous treatments of cilia and flagella
as line distributions of regularised forces~\citep{smith2009boundary} and
SRT, we generate a two-dimensional manifold surface mesh representing the ribbon
centreplane, and account for the finite thickness of the sheet via the
regularisation parameter $\epsilon$ of a surface distribution of regularised
forces.

The method is validated against analytical solutions for a translating plate
ellipsoid, and then compared with previous BEM~\citep{Keaveny2011} calculations
for the hydrodynamics of twisted helices and the resistance matrix of an
experimental ribbon-shaped artificial microswimmer.~\citep{Zhang2009} We then
demonstrate the flexibility of the method by calculating the flow around a
sedimenting double helix with a curved centreline, before finally considering
the swimming dynamics of a finite waving sheet~\citep{diller2014continuously} or
``magic carpet''. 

\section{Mathematical model}
We will consider microscale ribbons and sheets, for which the dynamics of the
surrounding fluid is well-modelled by the Stokes flow equations 
\begin{equation} 
    \mu\nabla^2\mathbf{u} -\nabla p = 0,
    \quad \nabla\cdot\mathbf{u}=0, 
    \label{eq:stokes_flow}
\end{equation} 
where $\mathbf{u}$ is the fluid velocity and $p$ the dynamic pressure.

To solve these equations~\eqref{eq:stokes_flow}, we employ the regularised
stokeslet boundary element method.~\citep{cortez2005method,smith2009boundary}
The velocity at a point $\mathbf{x}_0$ in the domain is given by integrals of
stokeslets $\mathbf{S}$ and stresslets $\mathbf{T}$ over the swimmer surface,
$S$,
\begin{align}
    \int_{V} u_j(\mathbf{x}) \phi_\epsilon(\mathbf{x} - 
    \mathbf{x}_0)\,\mathrm{d}V_{x} = \int_S
    &S_{ij}^\epsilon(\mathbf{x},\mathbf{x}_0)f_i(\mathbf{x})
    \nonumber \\
    &-u_i(\mathbf{x})T_{ijk}^\epsilon(\mathbf{x},\mathbf{x}_0)
    n_k(\mathbf{x})\,\mathrm{d}S_x,
    \label{eq:reg_bem}
\end{align}
for unknown surface tractions, $\mathbf{f}$, and surface velocity $\mathbf{u}$.
The left-hand side is a volume integral over the fluid domain of the velocity
multiplied by a `blob' regularisation of the Dirac $\delta$-function. 

For rigid body motions, or where the volume of a swimmer is constant over its
beat, we have the condition~\citep{Pozrikidis1992}
\begin{equation}
    \int_S \mathbf{u}(\mathbf{x})\cdot\mathbf{n}(\mathbf{x})\,\mathrm{d}S_x = 0.
    \label{eq:cond_slp}
\end{equation}
This condition also holds to a very good approximation for flexible yet inextensible swimmers provided
the thickness $h \ll w,l$, since the volume of such swimmers varies only very
slightly. Under condition~\eqref{eq:cond_slp}, the boundary integral
equation~\eqref{eq:reg_bem} can then be rewritten in terms of a modified force
density
\begin{equation}
    u_j(\mathbf{x}_0) = \int_S
    S_{ij}^\epsilon(\mathbf{x},\mathbf{x}_0)f_i(\mathbf{x})\,\mathrm{d}S_x,
    \label{eq:reg_bem_slp}
\end{equation}
which reduces computational complexity significantly. In our model, we will
impose velocities at the depth midplane and use the regularisation to account
for the finite thickness of the swimmer in a similar manner to Smith's treatment
of slender bodies;~\citep{smith2009boundary} we thus employ the ``single layer''
boundary integral equation~\eqref{eq:reg_bem_slp}. The regularisation function
for the blob driving forces is given by~\citep{cortez2005method}
\begin{equation}
    \phi_\epsilon(\mathbf{x} -
    \mathbf{x}_0) = \frac{15\epsilon^4}{8\pi r_\epsilon^7}, \quad r_\epsilon^2 =
    r^2 + \epsilon^2,
    \label{eq:blob_choice}
\end{equation}
where $r_i = (\mathbf{x} - \mathbf{x}_0)_i, r = |\mathbf{x} - \mathbf{x}_0|$ and
$\epsilon \ll 1$. For such a blob force, the regularised stokeslet is then
\begin{equation}
    S_{ij}^\epsilon(\mathbf{x},\mathbf{x}_0) = \frac{\delta_{ij}(r^2 +
    2\epsilon^2) + r_i r_j}{r_\epsilon^3}.
\end{equation}
We will discretise the boundary integral equation~\eqref{eq:reg_bem_slp} over
mesh of geometrically piecewise-quadratic triangles. The unknown tractions
$\mathbf{f}$ will be discretised as piecewise-constant $f_i[1],\ldots,f_i[N]$
over each element {$E[1],\ldots,E[N]$, where $E = E[1]\cup\ldots\cup E[N]$ is the
surface mesh of $N$ elements describing the ribbon or sheet. This discretisation
yields a matrix system of the form
\begin{equation}
    u_j(\mathbf{x}_0) = \sum_{n=1}^N f_i[n]\int_{E[n]}
    S_{ij}^\epsilon(\mathbf{x},\mathbf{x}_0)\,\mathrm{d}S_x,\quad \mathbf{x}_0
    \in E[m],
\end{equation}
where $\mathbf{x}_0$ is the centroid of element $E[m]$}, with $m = 1,\ldots,N$
and $i,j, = 1,2,3$. This yields $3N$ equations for the $3N$ unknown surface
tractions. Element integrals of the regularised stokeslets are performed using
adaptive Fekete quadrature. Code is implemented in matlab, and adapted from the
authors' previous work.~\citep{montenegro2015regularised} We will now proceed with a validation of this
method against previous analytical and numerical results.

\section{Validation}

\subsection{Plate ellipsoids}
We begin by considering the resistance of a plate ellipsoid to the $6$ principal
translations and rotations. The total drag $F$ on an ellipsoid with semi-axes
$\{\ell,w,h\}$, where without loss of generality $\ell \geq w \geq h$, translating at
speed $U$ in direction $\ell$ is given by~\cite{perrin1934radium}
\begin{equation}
    \frac{F}{\pi\mu U} = \frac{16}{\phi + \zeta_\ell \ell^2},
    \label{eq:el_drag}
\end{equation}
while the torque $T$ due to a rotation at rate $\Omega$ about $\ell$ is given by
\begin{equation}
    \frac{T}{\pi\mu\Omega} = \frac{16}{3}\frac{w^2 + h^2}{w^2\zeta_w +
    h^2\zeta_h},
\end{equation}
where
\begin{subequations}
    \label{eq:drag_int}
    \begin{align}
        \phi =& \int_0^\infty \frac{\mathrm{d}x}{\sqrt{(\ell^2 + x)(w^2 + x)(h^2
        + x)}},\\
        \zeta_\ell =& \int_0^\infty \frac{\mathrm{d}x}{(\ell^2+x)\sqrt{(\ell^2 + x)(w^2 + x)(h^2
        + x)}},\\
        \zeta_w =& \int_0^\infty \frac{\mathrm{d}x}{(w^2+x)\sqrt{(\ell^2 + x)(w^2 + x)(h^2
        + x)}},\\
        \zeta_h =& \int_0^\infty \frac{\mathrm{d}x}{(h^2+x)\sqrt{(\ell^2 + x)(w^2 + x)(h^2
        + x)}}.
    \end{align}
\end{subequations}
{The regularisation $\epsilon$ is chosen such that the thickness of
the semi-axis $h = \epsilon$, in the same way that $\epsilon$ is used as a proxy for
filament thickness in regularised stokeslet slender body
theory~\citep{smith2009boundary}. In practice, provided $\ell,w\gg h$, the drag
is fairly insensitive to this choice; substituting values of $\ell = 1, w = 0.1$
then $h = 0.001,0.0005$ into the above equations, we see less than a $0.3\%$
difference in all 6 components of the resistance matrix.}
Note that since we keep the regularisation constant over the entire surface, our
model in fact represents an elliptical disk of finite thickness. Nonetheless, for
very thin discs, we expect this discrepancy to be small.

\begin{figure}[h!]
    \begin{center}
	\includegraphics{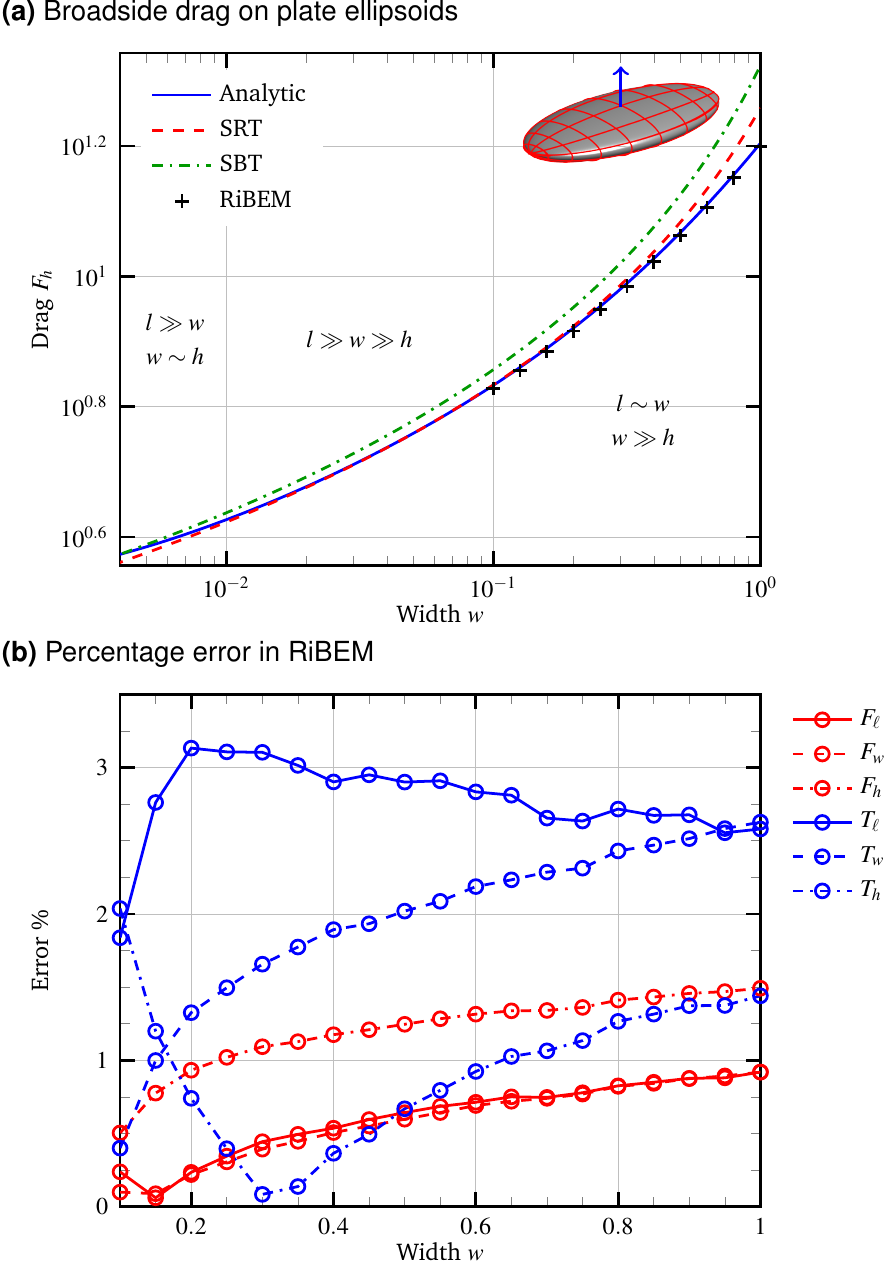}	
    \end{center}
\caption{(a) Comparing methods of evaluating the broadside drag (direction
demonstrated on the inset ellipsoid) $F_h$ on a plate
ellipsoid of length  $\ell = 1$, thickness $h = 0.005$, and varying width $w$.
RiBEM performs well until $w\sim0.1$ in a region roughly corresponding to
figure~\ref{fig:shape_classes}d, whereafter it because difficult to resolve
meshes. SRT is thereafter very accurate for this component in the region
corresponding to figure~\ref{fig:shape_classes}c, but is finally
superceded by SBT once width and thickness are comparable (the region of
figure~\ref{fig:shape_classes}b). At the bottom of the range, $w = h = 0.005$
and SBT gives the drag exactly. (b) The percentage
error in the principal components of the resistance matrix as calculated by
RiBEM for slightly thicker plate ellipsoids with $\ell = 1, h = 0.01$ and varying
$w$, showing good agreement with analytical results.}
\label{fig:plate_ellipsoid_error}
\end{figure}

Figure~\ref{fig:plate_ellipsoid_error}a shows a particular drag component, the
broadside drag $F_h$, on a thin plate ellipsoid of semi-axis length $\ell = 1$
and thickness $h = 0.005$ as a function of varying width $w$. The analytical
solution is compared to numerical solutions calculated via RiBEM, SRT, and SBT
assuming the radius of the filament is given by $w$. The RiBEM meshes
{for ellipse-based geometries} are
generated using routines adapted from DistMesh~\cite{persson2004simple}, with a desired uniform mesh element {edge} length
{dependent upon the area of the ellipse, $A = \pi\ell w$}, so that $E_l = \sqrt{\pi \ell
w/500}$.
{ Since an ideal mesh comprises equilateral triangles, which have an
area $A_t = \sqrt{3}E_l^2/4 \approx E_l^2/2$, this choice results
in uniform meshes with approximately $N=1000$ elements for each value of width
$w$.}
 
In the region $1\geq w \geq 0.3$, $\ell\sim w \gg h$, and the geometry is a sheet
as in figure~\ref{fig:shape_classes}d. RiBEM performs very favourably in
this r\'{e}gime, while the asymptotic solutions of SRT and SBT significantly
overestimate the drag. As $w$ decreases, we begin to separate all three length
scales $\ell\gg w \gg h$ (Fig.~\ref{fig:shape_classes}c), and for $0.3\geq w
\geq 0.1$, SRT has a similarly high accuracy to RiBEM, whereas SBT is still
significantly off. At around $w = 0.1$, $\ell \gg w$, and it becomes increasingly
difficult to numerically resolve the geometry using RiBEM, as retaining resolution
in the $w$ direction leads to a very fine mesh in the $\ell$ direction. This
results in large matrix systems with RiBEM, and so SRT becomes the appropriate
choice. At around $w = 0.01\sim h$, we no longer have separation of scales
between the width and thickness $\ell\gg w \sim h$ and the slender ribbon
assumption is no-longer valid. In this r\'{e}gime, (Fig.~\ref{fig:shape_classes}a,b)
SBT becomes the current most accurate choice, finally recovering the drag
exactly when $w = 0.005 = h$. Theses results demonstrate the need to choose the
correct method based on the separation of length scales in the geometry being
considered, as shown in figure~\ref{fig:shape_classes}. 

In figure~\ref{fig:plate_ellipsoid_error}b, we validate RiBEM for all $6$ values
of the resistance matrix for a thicker ellipsoid with $\ell = 1, h = 0.01,$ and
$w \in [1,0.1]$.  Finding all $6$ diagonal values of the resistance matrix with
approximately $N=1000$ elements took approximately $15$ seconds for each value
of $w$ on a Dell Optiplex 9020 desktop computer. Even for this relatively coarse
discretisation, the error in the mobility tensor is small, less than $3\%$ for
all components.  

\subsection{Helical ribbons} \label{sec:helicalribbons}

\begin{figure}[tb]
\begin{flushleft}
 \includegraphics{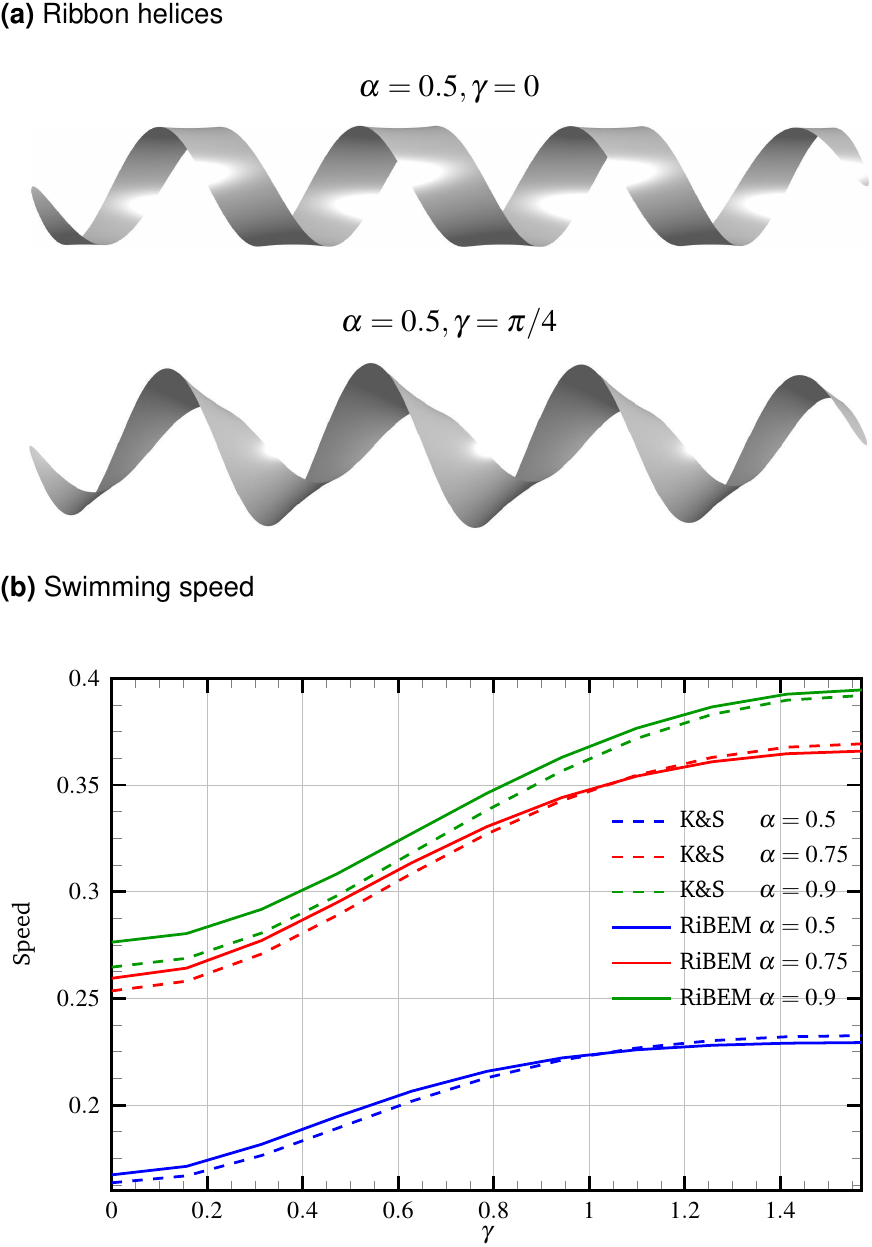}
\end{flushleft}
\caption{The dynamics of helical ribbons. (a) Ribbons given by the
parameterisation Eqs.~\eqref{eq:helixcent} and \eqref{eq:helixT} with $\alpha =
0.5$, and $\gamma = 0$ (left) and $\gamma = \pi/4$ (right). (b) The swimming
velocity in response to a unit torque about the helix axis for varying $\gamma$.
The dashed lines are the results of Keaveny and Shelly~\citep{Keaveny2011} for
ribbons with $h/w=1/4$, while the solid lines are the results from RiBEM for
ribbons with $\epsilon/w=1/5$. The slight discrepancy between these results is likely because
the Keaveny results have an elliptical cross-section, where as the RiBEM results
are for a flat cross-section.}
\label{fig:Keavenycomp}
\end{figure}

Recently, Keaveny and Shelley~\citep{Keaveny2011} explored how a helical
ribbon's configuration and aspect ratio, $h/w$, changed its propulsive
capability. This investigation used a boundary element method and considered
ribbons with an ellipsoidal cross-section and the parametrisation
\begin{align}
\mathbf{r}(s_{1}) =&\mbox{~}\left\{r_{h} \cos(k s_{1}),r_{h} \sin(k s_{1}),
\alpha s_{1}  \right\}, \label{eq:helixcent} \\
\mathbf{\hat{T}}(s_{1}) =&\mbox{~}\cos(\gamma) \mathbf{\hat{b}} -\sin(\gamma)
\mathbf{\hat{n}}. \label{eq:helixT}
\end{align}
Here, $\mathbf{r}(s_{1})$ is the helix centreline, $r_{h}$ is the helix radius,
$k = 4\pi$ is the helix wavenumber, and $\alpha$ is the cosine of the helix angle. The unit vector
$\mathbf{\hat{T}}(s_{1})$ points towards the major axis of
the ribbon cross-section, while $\gamma$ is the angle between the helix axis and
$\mathbf{\hat{T}}$. Finally, $\mathbf{\hat{n}}$ and $\mathbf{\hat{b}}$ are the normal
and binormal vectors to the helix centreline. Figure~\ref{fig:Keavenycomp}a
shows two example ribbons generated by this parameterisation. 

The thinnest of the ribbons considered by Keaveny and
Shelley~\citep{Keaveny2011}, with $h/w=1/4$, lie at the borderline where
traditional boundary element techniques can become computationally infeasible.
However, at these aspect ratios or finer, RiBEM is an effective alternative. 
Figure~\ref{fig:Keavenycomp}b shows the effect of varying $0 \leq \gamma \leq
\pi/2$ on configurations with $\alpha=0.5$, $0.75$ and $0.9$, with $h/w =
1/4$ for Keaveny and Shelley's results with an ellipsoidal cross-section
compared with RiBEM results for $\epsilon/w=1/5$ for a rectangular cross-section; this
aspect ratio is chosen such that the rectangular cross section $4hw \approx \pi
hw$ the ellipsoidal cross-section, as $\pi/16 = 0.196 \approx 1/5$.
These results are in very good agreement, {indicating that even for
relatively thick ribbons, this choice of $\epsilon=h$ provides a reasonable
proxy for finite thickness} with small deviations arising from the
difference in cross-sectional shape, which reduce for thinner ribbons. Indeed,
it should be noted that for manufacturing purposes, a rectangular cross-section
may in fact be preferable, as in the following case.

\subsection{A magnetic microhelix}

\begin{figure*}[tb]
\begin{flushleft}
\includegraphics{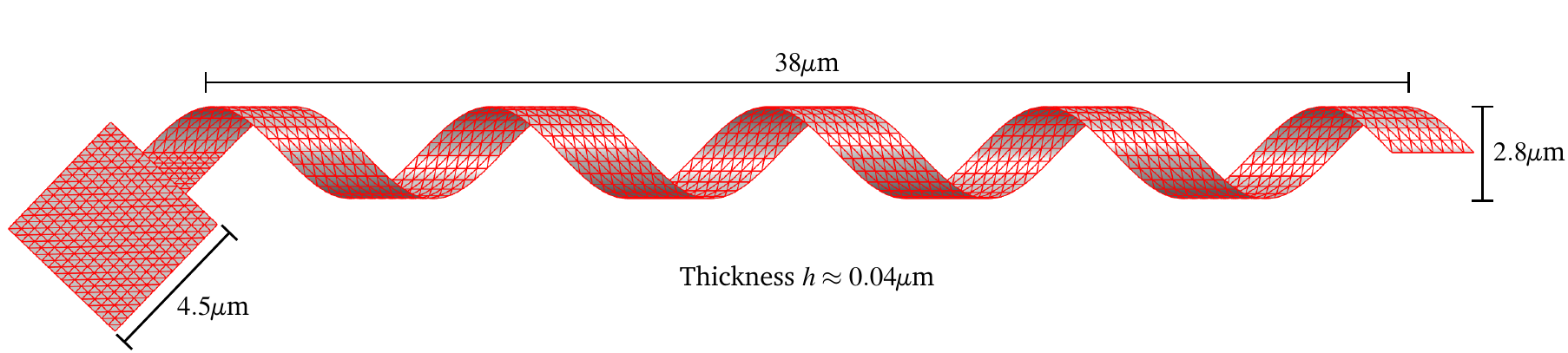}
\end{flushleft} 
\caption{Computational mesh of the magnetic helical microswimmer studied in
Ref.~\citep{Zhang2009} Note that this mesh differs
from previous numerical studies as the edges of the ribbon are not curved and
the head is not approximated by an ellipsoid, however these differences do not
have much of an effect on the resistance matrix, with the present study agreeing
with previous treatments using BEM for a significantly fatter mesh and SRT.}
\label{fig:nelson}
\end{figure*} 

Helical ribbons, similar to those considered above, have recently been used in
laboratory settings to create \textit{artificial bacterial flagella}
(ABF).~\citep{Zhang2009} These ABFs comprise a magnetic head rigidly attached to
a helical ribbon tail, and swim when driven by an external rotating magnetic
field. The swimming behaviour has in turn been used to experimentally determine
the resistance coefficients of such a swimmer,~\citep{Zhang2009} which have also
been analysed theoretically using SRT~\citep{Koens2016} and
BEM.~\citep{Keaveny2013} {The flexibility of RiBEM allows us
generate a realistic computational mesh of the experimental system
(Fig.~\ref{fig:nelson}). Indeed, provided that the geometric condition
$\ell/h,w/h \gtrsim 5$ holds, the approximate borderline as above between RiBEM and
BEM, RiBEM will be an effective method for the computational design and optimisation
of low Reynolds number artificial swimmers with ribbon or sheet-like
geometries.}

In the SRT study {of this ABF}, the ribbon was treated as
asymptotically thin, with the head modelled as a plate ellipsoid without
non-local hydrodynamic interactions. In the BEM study, the aspect ratio of $h/w
= 1/4$ overestimated the ribbon thickness by an order of magnitude. We now
attempt to use our model in order to resolve differences between the previous
studies and the experimental data.

The experimental ABF considered had a slender ribbon tail with an axial length
of $L=38$~$\mu$m, width $w = 1.8$~$\mu$m, and thickness $h = 42$~nm, and a
square magnetic head with dimensions of $4.5$~$\mu$m$\times 4.5$~$\mu$m $\times
200$~nm. The ribbon was helical, as a ribbon wrapped around a pencil, and had a
helix diameter of $2 r_{h} = 2.8$~$\mu$m immediately after fabrication. We
therefore describe this ribbon using Eq.~\eqref{eq:helixcent}, but with
$\mathbf{\hat{T}}= \mathbf{\hat{z}}$, where $\mathbf{\hat{z}}$ is the helix
axis. The mesh used for the RiBEM calculations, {generated using
simple custom routines,} is shown in figure~\ref{fig:nelson}.  This mesh is
closer to the experimental system than those used in previous
studies~\citep{Keaveny2013,Koens2016}, since it does not exhibit the curved
edges, which are not present in the experiment, found from the ribbon
parameterisation based upon Eqs.~\eqref{eq:helixcent} and \eqref{eq:helixT}.

Table~\ref{tab:helixswimmer} lists the experimentally-determined resistance
coefficients and the corresponding resistance coefficients of each model.
Significantly, we see that all independent theoretical models predict similar
values for the resistance coefficients, with slight variations likely due to the
details of each model. This result would seem to indicate that the discrepancy
was not caused by the approximations present in the previous SRT or BEM studies.
The drag coefficient $R_{a}$ is underpredicted in all models, while the
coefficient coupling axial rotation and translation $R_{b}$ is best captured
using the boundary element method or the RiBEM code. The rotational coefficient
$R_{c}$ is overestimated significantly by all three models. Koens and
Lauga~\citep{Koens2016} hypothesised that the curved edge of the SRT/Keaveny
parameterisation might be responsible for this discrepancy. However, since this
approximation is not present in our RiBEM model, the source of this discrepancy
remains unclear. 

\begin{table}[t]
\centering
\begin{tabular}{@{}l r r r r @{}} \toprule
& Exp  &  SRT   &   BEM &  RiBEM \\  \cmidrule{2-5}
$R_{a}$ ($10^{-7}$~N.s.m$^{-1}$)  & 1.5 & 1.04 &  0.937& 0.932\\
$R_{b}$  ($10^{-14}$~N.s) & -1.6  & -1.32 &  -1.63 & -1.47 \\
$R_{c}$  ($10^{-19}$~N.m.s)& 2.3 & 6.81&  10.1 & 9.91\\ \bottomrule
\end{tabular}
\caption{Hydrodynamic resistance coefficients for the ribbon microswimmer. 
Left: experimental measurements.~\citep{Zhang2009} Middle-Left: SRT results
assuming $2 r_{h}=2.8$~$\mu$m.~\citep{Koens2016} Middle-Right: the BEM model
swimmer results from Ref.~\cite{Keaveny2013} Right: RiBEM result assuming $2
r_{h}=2.8$~$\mu$m. The hydrodynamic resistance coefficient $R_{a}$ relates
the drag force experienced parallel to the helical axis from translation in the
same direction, whereas $R_{b}$ is the hydrodynamic force experienced parallel to the
helix axis from rotation around the helix axis, and $R_{c}$ is the hydrodynamic
torque experienced around the helical axis from rotation around said axis of
both the head and the helix.
}
\label{tab:helixswimmer}
\end{table}

\section{Results}

\subsection{A coiled double helix}

The helix above is an experimental realisation of the bacterial flagellum: a
microscale chiral structure that couples rotation with translation. In
the example above, the centreline of the helix is twisted about a central axis,
but the ribbon itself is not twisted about its centreline. Centreline twist is
another fundamental chirality which can couple rotation and translation, and is
observed in nature in, for instance, the double helix of of DNA. 

\begin{figure}[tb]
    \begin{center}
	\includegraphics{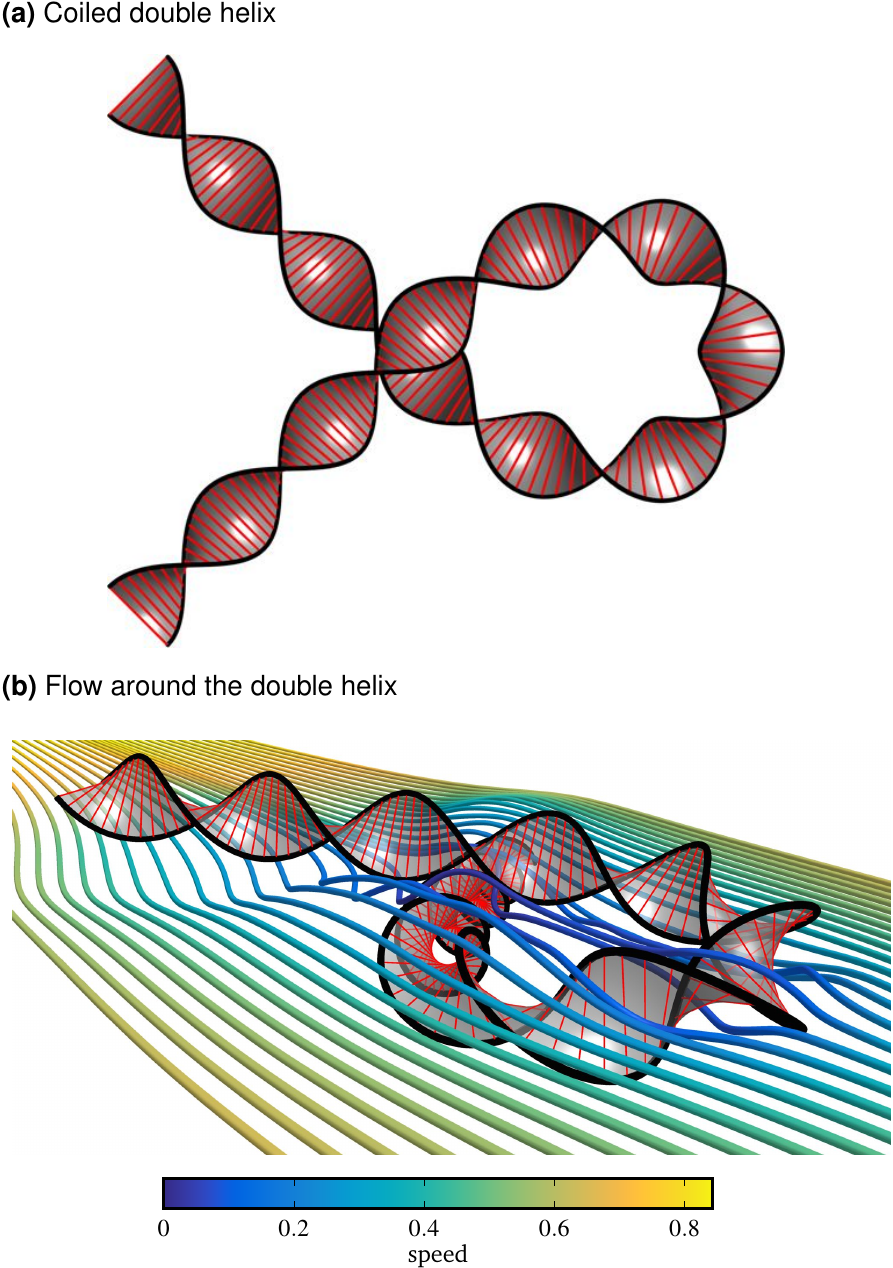}
    \end{center}
    \caption{(a) Plan view of a double helix with straight ends, coiled over
    into a half figure-of-eight. (b) Uniform flow past the double helix, showing
    streamlines that are deflected as though by an equivalent cylinder, and a few
streamlines that follow the twisting surface more closely.}
\label{fig:double_helix_mesh}
\end{figure}

In order to demonstrate the simplicity and flexibility of our method, we now
calculate the $6\times 6$ Grand Resistance Matrix (GRM) of a rigid, coiled double helix, and
the streamlines resulting from uniform flow past it. The centreline we prescribe
for the double helix is given by two straight lines, folding over into the
lemniscate of Bernoulli, given by the parametric equation
\begin{subequations}
\begin{align}
    x = & \frac{\sqrt{2}\cos t}{\sin^2 t + 1},  &x =&
    \frac{\pm(t\pm\pi/2)}{\sqrt{2}},\\
    y = & \frac{\sqrt{2}\sin t\cos t}{\sin^2 t + 1}, &y =&
    \frac{t\pm\pi/2}{\sqrt{2}},\\
    t \in\,& (-\pi/2,\pi/2), &t \in\,&[-\pi,-\pi/2],\quad t\in[\pi/2,\pi],
\end{align}
\end{subequations}
with $z = 0.1t,\ t\in[-\pi,\pi]$ . We generate constant twist along this
centreline, with $\mathbf{\hat{T}} = \cos(2\pi ks/L)\mathbf{\hat{b}} + \sin
(2\pi ks/L)\mathbf{\hat{n}}$ for centreline binormal $\mathbf{\hat{b}}$ and normal
$\mathbf{\hat{n}}$. Here, we denote the arclength of the centreline by $s$, which differs
from the parametric variable $t$, the length of the centreline $L$, and the
wavenumber $k = 6$. This parametrisation, with a width of $w = L/20$, produces the
folded double helix shown in figure~\ref{fig:double_helix_mesh}. 

Setting an arbitrary small thickness of $h = 0.01 \approx w/35$,
the resistance matrix for the double helix in this configuration, with a
resolution of $N = 3200$ elements, is given by
\begin{equation}
    \begin{bmatrix*}[c]
    13.0274& 0.0002& -0.0002&   -0.1688&   -0.0017&   -0.0017 \\
     0.0002& 14.0480& 0.2342 &   0.0004 &  -0.1690 &   0.2704  \\
    -0.0002& 0.2343& 16.3583&    0.0008&   -1.4422&    0.2710 \\
    -0.1688& 0.0005& 0.0009&  11.8006 &   0.0008 &   0.0005  \\
    -0.0015& -0.1693& -1.4422&    0.0004&   21.0341&    1.9619 \\
    -0.0018& 0.2704& 0.2706&    0.0005&    1.9623&   24.4865
\end{bmatrix*}
\nonumber
\end{equation}
It is worth noting that this is remarkably similar to the resistance matrix of a
cylinder extruded along the same centreline, capped by hemispheres, with radius
$w = L/40$ (i.e. half the radius of the ribbon)
\begin{equation}
    \begin{bmatrix*}[c]
        12.7157&   -0.0000&    0.0000&    0.1716&    0.0000&    0.0000 \\
        0.0000&   13.6380&    0.2477&    0.0000&    0.0891&   -0.3409  \\
        0.0000&    0.2476&   16.0148&   -0.0000&   -0.9322&   -0.2118  \\
        0.1716&    0.0000&    0.0000&   11.5256&    0.0000&    0.0000  \\
        -0.0000&    0.0891&   -0.9325&   -0.0000&   20.2099&    2.2285 \\
        -0.0000&   -0.3407&   -0.2120&    0.0000&    2.2281&   24.4529
\end{bmatrix*}
\nonumber
\end{equation}
with only some very small off-diagonal elements arising from the twisting of the
helix not present. This supports previous observations \citep{Koens2016} that
the GRM of a double helix is well-modelled by an ``average'' cylinder. We note
however that three of the terms in the DNA coupling sub-matrix have the opposite
sign to the cylinder case. This change in sign is due to the twisting of the
ribbon's surface creating a different flow pattern. However as these terms are
two orders of magnitude smaller than the largest terms in the GRM, only very
accurate experiments will be able to measure them.

The GRM only captures the trajectories of rigid objects, and does not account
for detailed flow structures at the ribbon surface or the spatially-dependent
force per unit area on the structure. Instantaneous flow streamlines for a
sedimenting double helix in the frame in which the body is stationary are shown
in figure~\ref{fig:double_helix_mesh}b, revealing intricate patterns and
twisting streamlines near the surface. {A particularly interesting
avenue of future research will be to couple this fully-resolved flow and
surface tractions to a constitutive model for the double helix itself.
\citet{omori2014numerical} modelled the deformation of a red blood cell, a
closed membrane travelling through a micropore, by coupling the standard
boundary element method for flow to a finite element solution for a thin
hyperelastic membrane representing the cell. Finite element shell computations
for open membranes could be coupled to RiBEM in an analogous way, indeed the
flexibility of regularised stokeslet methods makes them ideal for coupling with
elastic models for structures~\citep{olson2013modeling}.} 

\subsection{Taylor's magic carpet}

G. I. Taylor's model of a small-amplitude, infinite swimming
sheet~\cite{taylor1951analysis} is often used as an analytical means to study
microscale propulsion, and is increasingly being used to examine swimming in
non-Newtonian fluids.~\cite{velez2013waving,riley2015small} Furthermore, a
finite-sized analogue of Taylor's sheet has recently been developed
experimentally,~\cite{diller2014continuously} driven by an applied magnetic
field as a means of achieving controlled locomotion at small scales. 

\begin{figure}[h!]
    \begin{center}
	\includegraphics{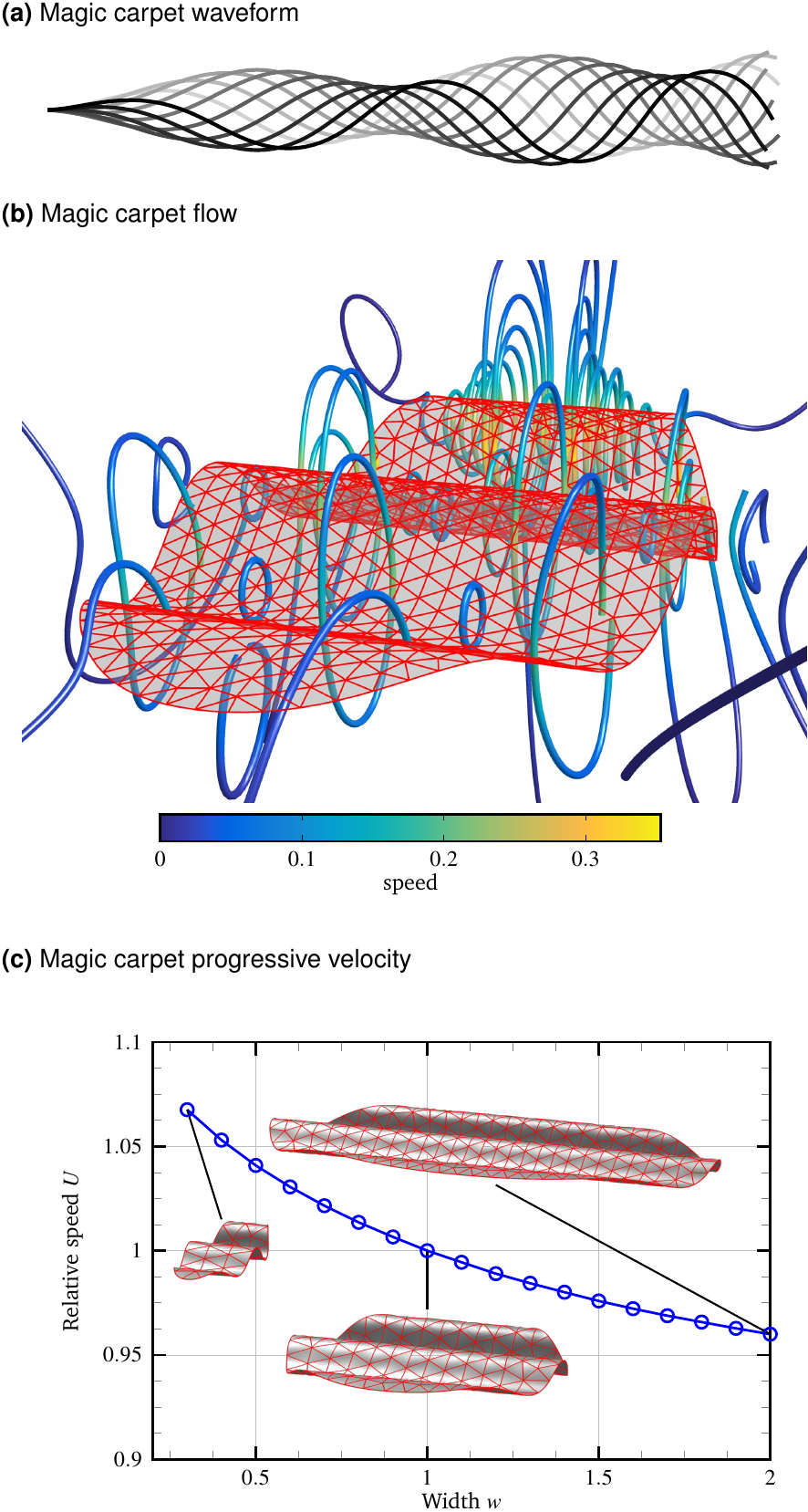}
     \end{center}
     \caption{(a) Grayscale time-lapse of the planar waveform of the magic carpet as defined by the
tangent angle~\eqref{eq:carpet_tangent}. (b) Flow streamlines driven by a magic carpet with $w = L/2, k =
         5\pi/2,$ and $A = \pi/3$ at time $t = 0$, demonstrating similar
         vortices to {those} observed in 2D and 3D analogues. (c) Swimming velocity
         of the magic carpet relative to a circular disk, demonstrating that for
         finite swimmers, slender filament configurations are faster than wide
     carpets exhibiting the same beat kinematics.}
\label{fig:magic_carpet}
\end{figure}

Our method allows us to model a range of plate microswimmers exhibiting
different dynamics. We consider an inextensible plate ellipsoid of length $L =
1$, thickness $h = 0.005$, and variable width $w$, propagating
bending waves along its length. Let $s = [0,1]$ be the length coordinate
relative to the head of the swimmer in an undeformed configuration, and $r =
[-w/2,w/2]$ the width coordinate. We model planar beating in the the length
$x$- and thickness $z$- plane, so that the tangent angle of the swimmer surface
in this plane is
\begin{equation}
\psi(s,t) = As^{1/2}\cos2\pi(ks - t),
\label{eq:carpet_tangent}
\end{equation}
giving body-frame surface coordinates
\begin{equation}
x = \int_0^1 \cos(\psi(s,t))\,\mathrm{d}s,\quad z = \int_0^1
\sin(\psi(s,t))\,\mathrm{d}s,\quad y = r,
\end{equation}
and body-frame velocity 
\begin{equation}
u = \int_0^1 -\dot{\psi}\sin(\psi(s,t))\,\mathrm{d}s,\quad w = \int_0^1
\dot{\psi}\cos(\psi(s,t))\,\mathrm{d}s,\quad v = 0.
\end{equation}
This parameterisation results in the beat pattern shown in figure~\ref{fig:magic_carpet}a.
The two unknown swimming $x$ and $z$ translational velocities and the rotational
velocity about the $y$-axis are found by enforcing the constraints that the
swimmer is force and torque free 
\begin{equation}
    \int_S \mathbf{f(\mathbf{x},t})\,\mathrm{d}S_x = \mathbf{0}, \quad \int_S
    \mathbf{x}\wedge\mathbf{f(\mathbf{x},t})\,\mathrm{d}S_x = \mathbf{0},
\end{equation}
and the sheet's mean progressive velocity determined from its laboratory frame
trajectory.~\citep{Koens2014}

An instantaneous configuration of a swimmer with $w = 0.5, k = 5\pi/2,$ and $A =
\pi/3$ at time $t = 0$ is shown in figure~\ref{fig:magic_carpet}b, together with
the resultant flow streamlines, in the laboratory frame. The streamlines reveal
an intricate pattern of recirculating vortices at the swimmer surface, as
observed both with filament swimmers and in two-dimensional (2D) infinite sheet
models; this intermediate model retains many of the flow features of both
limits.

We might also ask how the mean progressive velocity of such a swimmer varies
with width. Figure~\ref{fig:magic_carpet}c shows the progressive velocity of
this sheet swimmer as the width varies from $w = 0.3,\ldots,2$ relative to the
circular disk configuration. For the range considered, slender ribbons
exhibiting the same kinematics swim more quickly than broader swimmers.
{The hydrodynamic mechanism for this reduction for these parameters
can, somewhat crudely, be understood in the following manner. The propulsive
force that the sheet exerts on the fluid is approximately proportional to the
area of the ellipse, which scales with $w$. The drag in the swimming direction
can be approximated by $F_\ell$~\eqref{eq:el_drag}. The swimming velocity is
then such that drag and propulsion balance exactly, since the swimmer is
force-free. However, we can observe through simple numerical evaluation of the
integrals~\eqref{eq:drag_int} that the drag force~\eqref{eq:el_drag} increases
more rapidly with width than the surface area in this r\'{e}gime; thus for an
increase $dw$ in width, the increase in drag $dD > dF$ the increase in
propulsion, and the swimmer moves more slowly.}

{It is interesting to ask if the increase in velocity continues as
the width decreases further. By modelling the swimmer centreline by a line
distribution of 3D regularised stokeslets, we obtained a relative swimming speed
for the filament of $U = 1.032$. This result suggests that there may exist an
optimal cross-section which produces the fastest swimming speed in the nearly
filament r\'{e}gime (Fig.~\ref{fig:shape_classes}b) where there is currently no
non-local theory, providing further motivation for its development. However, it
is important to note that the regularised stokeslet slender body theory is a
different method for simulating fundamentally different geometries; whilst the
relative error between two RiBEM solutions for similar geometries will be small,
the absolute accuracy of RiBEM is around 3\%
(Fig.~\ref{fig:plate_ellipsoid_error}), and as such further study in the nearly
filament r\'{e}gime is required to verify the existence of this optimum.}

We again note that RiBEM would be highly suited for future coupling with a
shell theory that incorporates elastic bending in order to study actively
bending plate swimmers, in an analoguous manner to previous work with sperm-like
flagellar swimmers.~\cite{montenegro2015spermatozoa}

\section{Discussion}

We have demonstrated that the boundary element method with regularised
stokeslets is an effective means of calculating the dynamics of thin ribbons and
sheets, where the regularisation parameter $\epsilon$ is used as a proxy for the
sheet thickness. Tractions are not resolved across the thickness of the sheet,
but rather regularised stokeslets are distributed across a 2D manifold of the
ribbon's centreline. The method is accurate in regions where asymptotic theories
are not valid, and traditional boundary elements are impractical.

The method was validated against the analytical solution for the resistance
matrices of various plate ellipsoids as well as previous boundary element
studies of ribbon helices, and shown to be consistent with asymptotic and
boundary element studies of an experimental ribbon microswimmer. We then
applied the method to study the resistance and flow surrounding a coiled section
of double helix, finding that whilst the resistance matrix was very similar to
an equivalent cylinder, the flow streamlines showed an interesting structure which
would be important to resolve in fluid-structure interaction models. We finally
examined an inextensible, finite plate microswimmer or ``magic carpet'', finding
that for the parameters considered, making the plate wider slowed the swimmer
considerably.

The diverse geometries of natural and artificial bodies that interact with
microscale flows drives the need for bespoke numerical schemes that can solve
hydrodynamic problems efficiently and accurately. By applying a 
regularised stokeslet method, we have examined the dynamics of ribbons and
sheets that will be effective for a range of problems where the body thickness
is much smaller than the other dimensions, and may provide a means to couple
flows with plate mechanics to study fluid structure interaction problems such as
the sedimention of flexible pancake-like structures, or the unfolding of chiral
ribbons in ambient flow.

\subsection*{Acknowledgements}
TDM-J is supported by a Royal Commision for the Exhibition of 1851 Research Fellowship; LK is supported by the Cambridge Trusts, Cambridge Philosophical Society and the Cambridge hardship fund; El is supported in part by the European Union through a Marie Curie CIG Grant.

\balance



\providecommand*{\mcitethebibliography}{\thebibliography}
\csname @ifundefined\endcsname{endmcitethebibliography}
{\let\endmcitethebibliography\endthebibliography}{}

\end{document}